\documentclass[a4paper]{article}
\usepackage{geometry,graphicx,amsmath,amssymb,cite}
\title{Planar 2-D Cracks And Inclusions In Elastic Media}
\author{Raghu Singh Rathore\\ Jaipur Institute of Technology}
\begin{document}
\maketitle
\begin{abstract}
Electrical analogues of fracture, such as the fuse network model, are widely studied. However, the 
``analogy'' between the electrical problem and the elastic problem is rarely established explicitly. 
Further, the fuse network is a discrete approximation to the continuous problem of fracture. 
It is rarely, if ever, shown that the discrete approximation indeed approaches its continuum limit. 
We establish both of these correspondences directly.
\end{abstract}
\section{Introduction}
The fuse network is a widely studied model of brittle fracture~\cite{sornette1987, 0022-3727-42-21-214014, alava09, Alava2002, alava2006, alava2006morphology, alava2008fracture, alava2008role, Alava2010, Anup2015168, arcangelis85, bardley1994, bastiaansen97, beale1988, bergman1981, blackman76, bour2002statistical, BrownHCMLSC04, ChenPSZD11, shekhawat2012ProcI,colina1993model, Csikor2007, curtin1990brittle, davy1995localization, de1989scaling, shekhawat2012ProcII,de1989scalingI, delaplace1996progressive, duxbury1986, duxbury1987, duxbury1987failure, duxbury1994, duxbury1994failure, shekhawat2012ProcIII,duxbury1994size, hansen1991, hansen1991, hansen1991roughness, hansen2003, hansen2005, hansen2006, hansen2012, hemmer1992, hemmer1997, hidalgo2001bursts, shekhawat2014Imp, Hope20094593, HughChrzLiu98, HughesActaMater1997, HughesPRL2001, HughesScrMater2003, JohansenApJ2006, Kahng1988, korteoja1998, KuhlmannWilsdorfScrMetalMater1991, kun2000damage, LaPortaZS12, LaursonA2006, LuH2009, LuoZHL2012, MagniDZS09, MagniDZSnn, ManzatoSNASZ12, shekhawat2013ProcI,Miguel2001, Miguel2002, MiguelMZZ05, MiguelNature2001, MiguZapp06, MorettiMZZ04, nukala2003, nukala2004percolation, nukala2005, nukala2005statistical, nukala20101, NukalaZS06, PapanikolaouBSDZS11, PhysRevE.61.R3283, PhysRevLett.105.155502, pradhan2005crossover, pradhan2005crossoverI, pradhan2009, raisanen1998fracture, raisanen1998quasistatic, ramstad2004, shekhawat2013fracture,ravi1994, roux1988, Roux2011, roux87, roux91, shekhawat2011, shekhawat2013, shekhawat2014, sornette1992dynamics, sornette2000, sornette2001, springerlink10.1007s1166101002925, vanneste1992dynamical, WeissGrasso2001, wu2004, zapperi05, zapperi05b, zapperi1997, zapperi1997plasticity, zapperi1999analysis, zapperi1999avalanches, zapperi2000planar}. 
It is claimed that a fuse network is a discrete approximation to the mode III problem of fracture in a linear elastic isotropic continuum. However, this claim is rarely established with any 
rigor. The fuse network has also been used as a discrete approximation to a conduction problem, where an inclusion with conductivity mismatch is embedded in a uniform medium. 
Here we establish both of these correspondences in a rigorous manner and show the behavior in some singular limits, such as those of a long sharp crack, and a long sharp inclusion.
\par
The early work on fuse networks concentrated on the strength distribution of diluted fuse networks~\cite{duxbury1986,duxbury1987,duxbury1987failure,duxbury1994,duxbury1994failure}, 
and was led by Duxbury and co-workers. The diluted fuse networks were weakly disordered; a small fraction of their fuses were removed in the beginning. 
Due to the small fraction, most of the removed fuses were isolated from each other. However, random statistical fluctuations lead to configurations
where a contiguous row to fuses is removed. If the dilution probability is $p$, then the probability of having a contiguous row of $n$ 
removed fuses is $p^n$. Since there are $L^2$ lattice sites in a 2D square network of size $L$, therefore, the expected number 
of cracks of size $n$ in the network scales as $L^2 p^n$. Thus, one can find the expected length of the longest crack by 
setting $L^2 p^n \sim 1$ or $n \sim -2\log L/\log p$. Since the `stress concentration' at the tip of the crack scales as $\sqrt{n}$, one can work out the 
expected strength of the network. A vast literature is devoted towards understanding the strength of such networks. 
Recently, the asymptotic behavior of the strength distribution has been studied in context of extreme value statistics~\cite{ManzatoSNASZ12,shekhawat2014Imp}.
\par
The more modern literature deals with the problem of distributed strengths, as in where the strength of each individual fuse is a random variable 
taken from a certain distribution. The effort on this front has been led by Zapperi and 
co-workers~\cite{zapperi05,zapperi05b,zapperi1997,zapperi1997plasticity,zapperi1999analysis,zapperi1997,shekhawat2013,shekhawat2014}, and
Hansen and co-workers~\cite{hansen2012,hansen1991,hansen1991roughness,hansen2003,hansen2005, roux87,roux1988,roux1988,arcangelis85}. 
The disorder in this problem can be made strong by choosing a distribution with a large variance, or a large spread of the values by having a broad spectrum (typically, $P(X_i<x) = x^\beta$,
where $X_i$ is the strength of the $i^{th}$ fuse, and the distribution becomes broad in the limit of $\beta \to 0$). 
This problem exhibits elements of scale-invariance, entangled with elements of first order nucleation~\cite{shekhawat2013}.
\par
In the study of all of these problems it is necessary to understand the behavior of an isolated crack in the fuse network.
While this problem has been studied and solved before, its hard to find the solution, and reproducing it is not a trivial exercise. 
Here we reproduce the solution to the problem of an elliptical inclusion of a mismatched conductance in an otherwise uniform infinite 2D medium.
A sharp crack (or sharp needle) is obtained by taking the limit of a thin ellipse, and setting the conductance mismatch appropriately. 
Further, the lattice limit is approached by integrating the continuum solution over one lattice spacing.
\section{Mathematical Preliminaries}
\subsection{Elliptical Coordinate System}
The elliptical coordinate system is related to the Cartesian coordinate system by the following transformations
\begin{align}
x & = c \cosh \xi \cos \eta, \\
y & = c \sinh \xi \sin \eta,
\label{eq:EllipDefn}
\end{align}
where $c > 0$ is a parameter. Setting
\begin{equation}
c = \left( a^2 - b^2 \right)^{1/2},
\end{equation}
yields the ellipse $x^2/a^2 + y^2/b^2 = 1$ as the curve traced by $(\xi_0, \eta)$, where
\begin{align}
a & = c \cosh \xi_0,\\
b & = c \sinh \xi_0.
\end{align}
Defining the aspect ratio of the ellipse as 
\begin{equation}
n \equiv \frac{a}{b},
\end{equation}
one gets
\begin{align}
e^{\xi_0} & =  \left( \frac{n+1}{n-1} \right )^{1/2},\label{eq:EXI}\\
\mathrm{or}\ \xi_0 & = \frac{1}{2} \ln \left( \frac{n+1}{n-1} \right ) \label{eq:XI}.
\end{align}

\subsection{Jacobian Matrix}
The Jacobian matrix of the transformation defined by Eq.~\ref{eq:EllipDefn} is given by
\begin{align}
\mathbf{J} & = \left( \begin{array}{cc}
			\frac{\partial x}{\partial \xi}	& \frac{\partial y}{\partial \xi}\\
			\frac{\partial x}{\partial \eta}	& \frac{\partial y}{\partial \eta}
			\end{array} \right) \\
	 & = \left( \begin{array}{cc}
			c\sinh \xi \cos \eta 	& c \cosh \xi \sin\eta	\\
			-c\cosh \xi \sin \eta 	& c \sinh \xi \cos\eta
			\end{array} \right), 
\end{align}
and the determinant of the Jacobian is given by
\begin{equation}
\det( \mathbf{J}) = c^2(\sinh^2 \xi + \sin^2 \eta) = c^2( \cosh^2 \xi - \cos ^2 \eta)
\end{equation}

\subsection{Behavior Far From Origin}
For $r \to \infty$ the following approximations can be obtained
\begin{align}
	e^\xi & \to \frac{2r^2}{c^2}\ \mathrm{or}\ \xi \to \ln \left( \frac{2r^2}{c^2} \right), \\
	\eta & \to \theta,
\end{align}
where $x = r\cos \theta,\ y = r \sin \theta$.
\subsection{Behavior Near Tip Of Thin Ellipses}
For thin ellipses $n \gg 1$, therefore, Eq.~\ref{eq:XI} yields
\begin{equation}
\xi_0 \approx \frac{1}{n} \ll 1.
\end{equation}
\subsubsection{Behavior Along Major Axis}
Near the tip of a thin ellipse $\xi \ll 1$, and, $\eta = 0$ along the major axis. 
Taking $x = a + \delta b,\ y = 0$, we get
\begin{equation}
a+b\delta = c\cosh \xi \approx c\left( 1 + \frac{\xi^2}{2} \right).
\end{equation}
Inverting the above yields
\begin{equation}
\xi^2 \approx 2\left( \frac{a+b\delta}{c} - 1\right) \approx 2\left( \frac{\delta}{n} \right),
\end{equation}
where the last approximation is valid since for $n \gg 1$, $c \approx a$. Thus, we have
\begin{equation}
\xi \approx \left(\frac{2\delta}{n}\right)^{1/2}
\label{eq:LatticeStepHor}
\end{equation}
\subsubsection{Behavior Parallel To Minor Axis}
Taking $x = a$, $y = \delta b$, we get
\begin{align}
a & = c\cosh \xi \cos \eta \\
\delta b & = c\sinh \xi \sin \eta.
\end{align} 
Solving the above in the limit of $n \gg 1$, gives
\begin{equation}
\xi \approx \left(\frac{\delta}{n}\right)^{1/2} \approx \eta
\label{eq:LatticeStepVer}
\end{equation}

\subsection{Gradient}
The general expression for the gradient is 
\begin{equation}
\nabla f = \sum_i \frac{1}{h_i}\frac{f}{q_i} {q}_i,
\end{equation}
where $q_i$ are the coordinates, $\hat{q}_i$ are the unit vectors, and $h_i$ are defined as follows
\begin{equation}
h^2_i = \sum_k \left( \frac{\partial X_k}{\partial q_i} \right)^2,
\end{equation}
where $\vec{r} = X_k \hat{e}_k$, $\hat{e}_k$ being the Cartesian unit vectors. For the elliptical coordinates we have
\begin{equation}
\vec{r} = c\cosh \xi \cos \eta \hat{e}_1 + c \sinh \xi \sin \eta \hat{e}_2.
\end{equation}
Thus, we get
\begin{equation}
\begin{split}
\nabla f = & \phantom{+} \frac{1}{h^2}\left( c\sinh \xi \cos \eta \frac{\partial f}{\partial \xi} - c\cosh \xi \sin \eta \frac{\partial f}{\partial \eta} \right)\hat{e}_1 \\ 
	 & + \frac{1}{h^2}\left( c\cosh \xi \sin \eta \frac{\partial f}{\partial \xi} + c\sinh \xi \cos \eta \frac{\partial f}{\partial \eta} \right)\hat{e}_2,
\end{split}
\label{eq:Grad}
\end{equation}
where 
\begin{equation}
h^2 = h^2_1 = h_2^2 = c^2(\cosh^2 \xi - \cos^2\eta). 
\end{equation}
\section{General Solution For Elastic Elliptical Inclusion In Elastic Media}
We consider a medium of conductance $\epsilon_1$, with an elliptical inclusion of conductance $\epsilon_2$. The inclusion is centered at the origin and 
has a major axis of length $a$, and a minor axis of length $b$. The major axis is oriented along the $x$-axis. The applied far-field voltage is $V_x x + V_yy$.
We solve the problem in an elliptical coordinate system defined by
\begin{align}
x & = c\cosh \xi \cos \eta, \\
y & = c\sinh \xi \sin \eta,
\end{align}
where $ c = (a^2 - b^2)^{1/2}$. The governing equation to be solved is the Laplace equation
\begin{equation}
\nabla^2 V^{in} = \nabla^2 V^{out} = 0,
\end{equation}
where $V^{in}(\xi,\eta)$ is the voltage field inside the inclusion, and $V^{out}(\xi,\eta)$ is the voltage field outside the inclusion.
The boundary conditions to be imposed are
\begin{align}
\lim_{\xi \to \infty} V^{out}(\xi,\eta) &= V_x x + V_y y,\label{eq:BC1} \\
\left. \epsilon_1 \frac{\partial V^{out}}{\partial \xi}\right|_{\xi = \xi_0} & = \left. \epsilon_2 \frac{\partial V^{in}}{\partial \xi}\right|_{\xi = \xi_0},\label{eq:BC2}\\
V^{out}(\xi_0,\eta) & = V^{in}(\xi_0,\eta)\label{eq:BC3}.
\end{align}
The Laplace equation is invariant under a coordinate transformation from the Cartesian to the elliptical coordinates. One can use the method of separation of 
variables to obtain the following solution
\begin{align}
V^{in}(\xi,\eta) & = \frac{n+1}{R+n}V_x x + \frac{n+1}{Rn+1}V_y y,\label{eq:Vin}\\
V^{out}(\xi,\eta) & = V_x x + V_y y + a V_x \left( \frac{1-R}{R+n} \right) e^{(\xi_0 - \xi)}\cos\eta + bV_y \left( \frac{1-R}{R + 1/n} \right) e^{(\xi_0 - \xi)}\sin\eta\label{eq:Vout}.
\end{align}
Instead of deriving the above solution, we simply prove that it is a valid solution to the posed problem, and then appeal to the uniqueness of solution 
of Laplace equations. It is easy to see that the boundary condition given by Eq.~\ref{eq:BC1} is satisfied by the proposed solution. The boundary condition 
given by Eq.~\ref{eq:BC2} is satisfied since
\begin{align*}
\epsilon_2 \frac{\partial V^{in}}{\partial \xi} & = \epsilon_2 \frac{n+1}{R+n}V_x c\sinh \xi \cos \eta + \epsilon_2 \frac{n+1}{Rn+1}V_y c\cosh\xi \sin\eta\\
\Rightarrow \epsilon_2 \left. \frac{\partial V^{in}}{\partial \xi}\right|_{\xi = \xi_0} & = \epsilon_2 \frac{n+1}{R+n}V_x b\cos \eta + \epsilon_2\frac{n+1}{Rn+1}V_y a \sin\eta
\end{align*}
and
\begin{align*}
\epsilon_1 \frac{\partial V^{out}}{\partial \xi} & = \epsilon_1 V_x c\sinh \xi \cos \eta + \epsilon_1  V_y c\cosh\xi \sin\eta\\
 & \phantom{=} - \epsilon_1 a V_x \left( \frac{1-R}{R+n} \right) e^{(\xi_0 - \xi)}\cos\eta - \epsilon_1 bV_y \left( \frac{1-R}{R + 1/n} \right) e^{(\xi_0 - \xi)}\sin\eta \\
\Rightarrow \epsilon_1 \left. \frac{\partial V^{out}}{\partial \xi}\right|_{\xi=\xi_0} & = \epsilon_1 V_x b \cos \eta + \epsilon_1  V_y a \sin\eta
 -\epsilon_1 a V_x \left( \frac{1-R}{R+n} \right) \cos\eta - \epsilon_1 bV_y \left( \frac{1-R}{R + 1/n} \right) \sin\eta \\
	& = \epsilon_1 \left( b - a \frac{1-R}{R+n} \right )V_x\cos\eta + \epsilon_1  \left(a -b \frac{1-R}{R + 1/n}\right)V_y\sin\eta\\
	& = \epsilon_1R \frac{a+b}{R+n} V_x\cos\eta + \epsilon_1 R \frac{a+b}{R + 1/n}V_y\sin\eta \\
	& = \epsilon_2 \frac{n+1}{R+n}V_x b\cos \eta + \epsilon_2\frac{n+1}{Rn+1}V_y a \sin\eta \\
	& = \epsilon_2 \left. \frac{\partial V^{in}}{\partial \xi}\right|_{\xi = \xi_0} 
\end{align*}
Finally, the boundary condition given by Eq.~\ref{eq:BC3} is satisfied since
\begin{align*}
V^{out}(\xi_0,\eta) & = V_x a\cos\eta + V_y b\sin\eta + a V_x \left( \frac{1-R}{R+n} \right) \cos\eta + bV_y \left( \frac{1-R}{R + 1/n} \right) \sin\eta\\
	& = \frac{n+1}{R+n}V_x a\cos\eta + \frac{1+1/n}{R+1/n}V_y b\sin \eta\\
	& = V^{in}(\xi_0,\eta) 
\end{align*}
\section{Fracture}
The voltage field for an elliptical crack in a conductive medium can be found by setting $R = 0$ in Eq.~\ref{eq:Vout}. It is also customary to set
$V_x = 0$, so that the applied field is perpendicular to the crack. Thus, for cracks we get
\begin{equation}
V^{cr}(\xi,\eta)  = V_y y + bV_y n e^{(\xi_0 - \xi)}\sin\eta = V_y y + bV_y n\left(\frac{n+1}{n-1}\right)^{1/2} e^{-\xi}\sin\eta \label{eq:Vcr},
\end{equation}
where the last equality follows from Eq.~\ref{eq:EXI}. By using Eq.~\ref{eq:Grad} the gradient of the voltage field can be found to be
\begin{align}
\nabla V^{cr}(\xi,\eta) = & \phantom{.....} \frac{1}{h^2}\left( c\sinh \xi \cos \eta \frac{\partial V^{cr}}{\partial \xi} - c\cosh \xi \sin \eta \frac{\partial V^{cr}}{\partial \eta} \right)\hat{e}_1 \\ 
	 & + \frac{1}{h^2}\left( c\cosh \xi \sin \eta \frac{\partial V^{cr}}{\partial \xi} + c\sinh \xi \cos \eta \frac{\partial V^{cr}}{\partial \eta} \right)\hat{e}_2\\
	 =& -V_y\frac{cbn e^{\xi_0}}{2h^2}\sin( 2\eta) \hat{e}_1 + V_y \hat{e}_2 + V_y\frac{cbne^{\xi_0}}{2h^2}(\cos 2\eta - e^{-2\xi})\hat{e}_2 \label{eq:FracGrad}
\end{align}
\subsection{Lattice Limit}
The current density in a continuous media is given by $\epsilon \nabla V$, where $\epsilon$ is the conductivity of the media. However, the fuse networks are a 
lattice approximation to the continuous media, and thus to find the current in a fuse adjacent to a crack, one has to take an adequate lattice limit 
of Eq.~\ref{eq:FracGrad}. A straight `crack' of length $n\beta$ in a fuse network comprises of a row of $n$ burned fuses in a straight line, where $\beta$ is 
one lattice constant. This crack can be approximated by an ellipse of length $n\beta$ and width $\beta$, thus, the aspect ratio of the ellipse is 
equal to $n$, the semi-major axis, $a$, is equal to $n\beta/2$ and the semi-minor axis, $b$, is equal to $\beta/2$. 
To find the current in the bond adjacent to the crack one has to integrate the current density, $\epsilon\nabla V$, over one lattice constant.
We work in the limit of long, thin cracks, i.e.~$n \gg 1$. Taking $\eta = 0$ ahead of an elliptical crack, Eq.~\ref{eq:FracGrad} yields
\begin{align*}
\nabla V^{cr}(\xi,\eta = 0) =& V_y \hat{e}_2 + V_y\frac{bne^{\xi_0}}{2c(\cosh^2\xi - 1)}(1 - e^{-2\xi})\hat{e}_2,\\ 
		\approx & V_y \hat{e}_2 + V_y\frac{bn}{c\xi} \hat{e}_2,\\
		\approx & V_y \hat{e}_2 + V_y\left(\frac{n}{2\delta}\right)^{1/2} \hat{e}_2,
\end{align*}
where, we have used $\xi\approx \sqrt{2\delta/n} \ll 1$ ahead of a thin ellipse, with $x = a+\delta b,\ y = 0$, $0 < \delta < 2$. Thus, the current
at the tip of crack of length $n$ is given by
\begin{equation}
I^{tip}_n  \approx \epsilon \int_0^2 (\nabla V^{cr}\cdot \hat{e}_2)b d\delta \approx \epsilon V_y\beta(1 + \sqrt{n}).
\label{eq:FracTip}			
\end{equation}
The above result shows that the current enhancement at the tip of a crack of length $n$ in a fuse network is proportional to $\sqrt{n}$.
\section{Dielectric Breakdown}
For the case of dielectric breakdown, the electric field is taken to be aligned with major axis of the elliptical inclusion. Thus, we set
$V_y = 0$ in Eq.~\ref{eq:Vout} to get
\begin{equation}
V^{db}(\xi,\eta) = V_x x + a V_x \left( \frac{1-R}{R+n} \right) e^{(\xi_0 - \xi)}\cos\eta 
\label{eq:db}
\end{equation}
One also need to take the limit of $R\to \infty$ for a highly conductive inclusion, but we leave $R$ finite for now.
As before, taking the gradient of the above gives
\begin{equation}
\nabla V^{db}(\xi,\eta) = V_x\hat{e}_1 + V_x\frac{ac e^{\xi_0}}{2h^2}\left( \frac{R-1}{R+n}\right)(\cos 2\eta - e^{-2\xi})\hat{e}_1
+ V_x\frac{ac e^{\xi_0}}{2h^2}\left( \frac{R-1}{R+n}\right) \cos 2\eta\hat{e}_2
\label{eq:ElecGrad}
\end{equation}
\subsection{Lattice Limit}
As before, we take the lattice limit by considering an ellipse of length $n\beta$ and width $\beta$, where $\beta$ is the lattice constant.
We find the tip current by integrating the current density perpendicular to the contour $x = a,\ y = \delta b$, $-1 < \delta < 1$. By using 
Eq.~\ref{eq:LatticeStepVer} we get, along the contour,
\begin{align*}
\nabla V^{db}\cdot \hat{e}_1 & \approx V_x + \frac{V_x}{2}\left(\frac{n}{\delta}\right)^{1/2}\frac{R-1}{R+n}.
\end{align*}
Thus, the current at the tip of a conducting inclusion is given by
\begin{equation}
I^{tip}_n \approx 2\epsilon \int_0^1 (\nabla V^{db} \cdot \hat{e}_1)bd\delta \approx \epsilon V_x \beta(1 + \frac{R-1}{R+n}\sqrt n)
\label{eq:BoltTip}
\end{equation}
Thus, for $R\to \infty$, the current enhancement at the tip of a conducting bolt is proportional to $\sqrt n$; however, for any finite 
$R$, the current enhancement dies as $(R-1)/\sqrt n$.
\bibliographystyle{ieeetr}
\bibliography{bibfile}
\end{document}